# Spin-dynamics of the low-dimensional magnet $(CH_3)_2NH_2CuCl_3$

Matthew B. Stone[a], Wei Tian[b], Garrett E. Granroth[a], Mark D. Lumsden[a], J-H. Chung[c], David G. Mandrus[a], Stephen E. Nagler[a]

[a] Oak Ridge National Laboratory, Oak Ridge, TN 37831, USA

[b] Department of Physics and Astronomy, The University of Tennessee, Knoxville, TN 37996, USA

[c] NIST Center for Neutron Research, Gaithersburg, MD 20899, USA

**Abstract**

Dimethylammonium copper (II) chloride (also known as $DMACuCl_3$ or MCCL) is a low dimensional $S=1/2$ quantum spin system proposed to be an alternating ferro-antiferromagnetic chain with similar magnitude ferromagnetic (FM) and antiferromagnetic (AFM) exchange interactions. Subsequently, it was shown that the existing bulk measurements could be adequately modeled by considering $DMACuCl_3$ as independent AFM and FM dimer spin pairs. We present here new inelastic neutron scattering measurements of the spin-excitations in single crystals of $DMACuCl_3$. These results show significant quasi-one-dimensional coupling, however the magnetic excitations do not propagate along the expected direction. We observe a band of excitations with a gap of $\Delta = 0.95$ meV and a bandwidth of 0.82 meV.

*Keywords:* Antiferromagnet; S=1/2; low-dimensional, organo-metallic

## 1. Introduction

Quasi-one-dimensional (1d) $S = 1/2$ alternating Heisenberg chains continue to attract attention theoretically and experimentally as model magnetic systems. Particularly interesting are systems with comparable magnitude exchange but alternating between ferromagnetic and antiferromagnetic interactions. The material we examine, $DMACuCl_3$, has been considered to be within this category. $DMACuCl_3$ was originally studied in the 1930's [1]. Willett proposed that the system was quasi-one dimensional (1d), with $S = 1/2$ $Cu^{2+}$ ions coupled via Cu-halide-Cu bridges to form magnetic chains along the crystalline $a$-axis [2]. Magnetic susceptibility measurements and structure calculations led to the classification of $DMACuCl_3$ as an alternating sign exchange chain with antiferromagnetic (AFM) exchange $J_{AFM} = 1.1$ meV and ferromagnetic (FM) exchange $J_{FM} = 1.3$ meV alternating along the chain axis [3,4]. However, the nature of the magnetic behavior in this system has only been postulated based upon the structure and thermodynamic measurements that can often be well described by a number of theoretical model calculations [5]. We provide here the first spectroscopic measurements of the magnetic excitation spectrum of $DMACuCl_3$ via inelastic neutron scattering measurements. We find that the magnetic coupling is not along the proposed $a$-axis, rather the primary quasi-1d direction is along the crystalline $b$-axis.

## 2. Results and Discussion

$DMACuCl_3$ crystallizes in the monoclinic structure (space group I2/a) with room temperature lattice constants $a = 11.97$ Å, $b = 8.6258$ Å, and $c = 14.34$ Å, and $\beta = 97.47°$ [2]. The crystal structure consists of Cu-Cl-Cu bonded chains along the crystalline $a$-axis with Cu-Cl-Cl-Cu contacts along the crystalline $b$-axis. These Cu-halide planes are separated from one another along the longer $c$-axis by methyl groups.

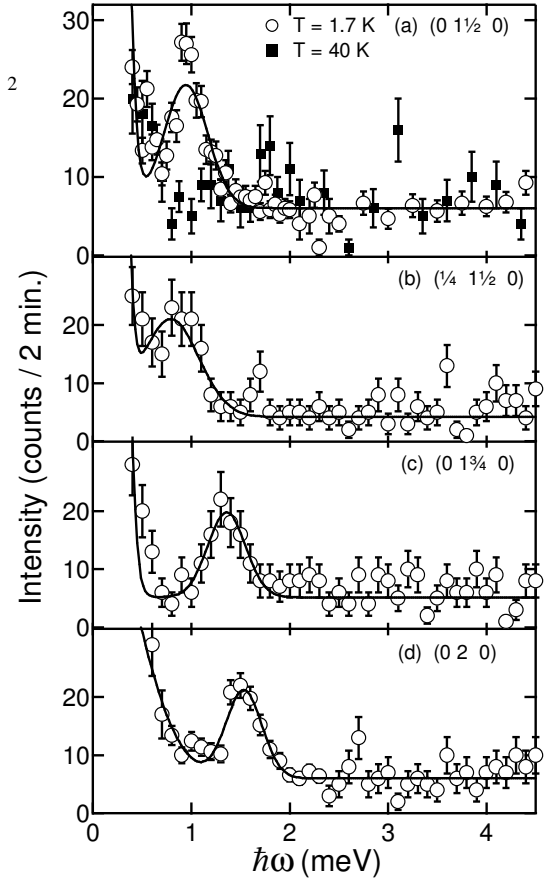

Fig. 1. Constant wave-vector measurements of DMACuCl$_3$ at representative points of the excitation spectrum. Panel (a) compares a $T = 1.7$ K measurement with a $T = 40$ K measurement. Panel (a) is on a different vertical scale than (b), (c) and (d). Solid lines are fits to Gaussian lineshapes.

Deuterated single crystal samples were prepared through slow isothermal evaporation of stoichiometric solutions of CuCl$_2$·2(H$_2$O) and (CH$_3$)$_2$NH$_2$Cl dissolved in D$_2$O [2]. The sample for our measurements consisted of a single $m = 2.23$ g crystal oriented in the ($hk0$) scattering plane. Sample environment consisted of a pumped He$^4$ cryostat. Inelastic neutron scattering measurements were performed using the cold-guide based triple axis SPINS spectrometer at the NIST neutron scattering facility in Gaithersburg, Maryland. 80' collimation was used before and after the sample. A fixed final energy of 5 meV was used throughout the measurement with a flat (non-focusing) 6.3 cm wide pyrolytic graphite (PG (002)) analyzer crystal. A liquid nitrogen cooled Beryllium filter was in place in front of the analyzer. Constant wave-vector scans were performed throughout the ($hk0$) scattering plane counting for a fixed number of beam-monitor counts, using a low-efficiency beam monitor placed in front of the sample.

Figure 1 depicts constant wave-vector scans performed at (0 1½ 0), (¼ 1½ 0), (0 1¾ 0) and (0 2 0). The temperature dependence of the (0 1½ 0) excitation is shown in Fig. 1(a). At $T = 40$ K, the low-temperature peak observed at $\hbar\omega = 0.9$ meV is significantly reduced in intensity indicating that the low-temperature excitations are magnetic in origin. Comparisons between Fig. 1(a) and 1(b) indicate a lack of dispersion for the magnetic excitation along the ($h$00) direction. The peaks shown in Figs. 1(a), (c) and (d) indicate a single dispersive mode propagating along the (0$k$0) direction.

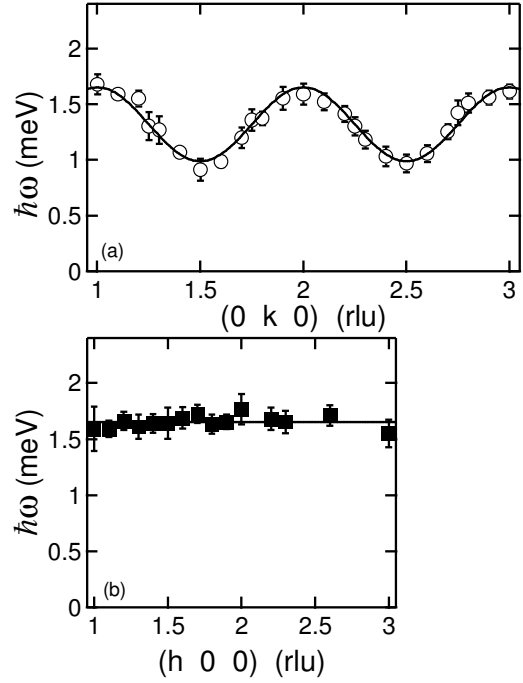

Fig. 2. $T = 1.7$ K excitation spectrum of DMACuCl$_3$ along the (a) (0$k$0) and (b) ($h$00) directions. Horizontal scaling reflects the differences in length of the reciprocal lattice vectors $a^*$ and $b^*$. Points represent peak locations of constant wave-vector scans using a Gaussian approximation for the lineshape. Solid line is a comparison to the dispersion discussed in the text.

Figure 2 depicts the magnetic peak positions for a series of constant wave-vector scans along the (a) (0$k$0) and (b) ($h$00) directions. There is no dispersion of the single mode along the ($h$00) direction; however there is significant dispersion along the (0$k$0) direction. These results clearly indicate that the quasi-1-d axis is along the (0$k$0) direction corresponding to magnetic exchange along the crystalline $b$-axis, contrary to the proposed quasi-1d axis. The mode has a spin-gap of $\Delta = 0.95$ meV and a bandwidth of 0.82 meV. Constant wave-vector scans were also measured along the ($h\zeta0$) direction for $\zeta = $ ¼, ½, ¾, 1 and 1½; none of these indicated any dispersion along the $h$ reciprocal lattice vector beyond instrumental resolution.

The single particle dispersion observed is commensurate with the underlying lattice and is of the form

$$\hbar\omega(\mathbf{Q}) = A + B/2 \, \cos(2\pi k). \qquad (1)$$

Comparing this to the results shown in Fig. 2, we determine $A = 1.32(1)$ meV and $B = 0.66(3)$ meV. Eqn. 1 corresponds both to the approximate dispersion associated with a Heisenberg two-leg spin ladder with a strong rung bond [6] and to the dispersion of an alternating Heisenberg antiferromagnetic chain [7] illustrating that a unique model of the magnetic behavior is not available based solely on the dispersion. The intensity modulation along both the (0$k$0) and ($h$00) directions varies slowly over the wide range of wave-vectors measured. The observed intensity is consistent with spin dimerization but more detailed measurements may be required to formulate a definitive model for the magnetic Hamiltonian. The crystal structure indicates that there is more than one candidate for the spin-dimer; however, interdimer interactions are clearly present along the crystalline $b$-axis leading to the dispersion observed in Fig. 2(a). We also note that a high temperature, T ≈ 280 K, structural phase transition occurs in DMACuCl$_3$ [8]. This phase transition likely affects the nature of the Cu-Cl-Cu and Cu-Cl-Cl-Cu spin exchange.

We observe no higher energy excitations for energy transfers as large as $\hbar\omega$ = 5 meV. This is significant given the proposed alternating exchange model for DMACuCl$_3$. Although the theory of the alternating bond (FM-AFM) spin chain in the absence of any anisotropy predicts a gapped excitation for the entire range of $|J_{FM}/J_{AFM}|$ values [9,10], finite chain calculations also predict a higher energy continuum of excitations for a range of $|J_{FM}/J_{AFM}|$ [11,12]. The prior proposed model of magnetic interactions in DMACuCl$_3$ was in the vicinity of $|J_{FM}/J_{AFM}| \approx 1$; at this point a continuum of excitations exists at the zone boundaries, at the maximum in the single particle dispersion, with a second gap to higher energy excitations at the zone center [12]. Our current results reveal neither of these effects, further invalidating the FM-AFM spin chain model in DMACuCl$_3$.

## 3. Conclusion

We have performed inelastic neutron scattering measurements on the organo-metalic magnet DMACuCl$_3$. These measurements reveal that this system is *not* a quasi-1d FM-AFM spin chain along the crystalline $a$-axis as originally proposed. Rather we find the significant dispersive direction to be the (0$k$0) direction, indicating that the $b$-axis is the 1d magnetic axis. Preliminary thermal neutron scattering measurements reveal no significant dispersion along the $c$-axis as expected based upon the large $c$-axis lattice constant and intervening ligands between the Cu-halide groups within the $ab$ plane. The amount of dispersion observed for the single-particle excitation also complicates the interpretation of thermodynamic models of this system based upon isolated ferromagnetic and antiferromagnetic spin dimers [4,13]. The wave-vector distribution of scattering intensity and the dispersion indicates that the magnetism in DMACuCl$_3$ is dominated by a spin-dimer that is weakly coupled along the $b$-axis. However to completely elucidate the underlying Hamiltonian will require further analysis and measurements.

## ACKNOWLEDGEMENTS

Oak Ridge National laboratory is managed by UT-Battelle, LLC, for DOE under DE-AC05-00OR22725. Work at SPINS supported by the National Science Foundation under Agreement No. DMR-9986442, and the National Institute of Standards and Technology, U. S. Department of Commerce. We acknowledge valuable discussions with M. W. Meisel.